# Atmospheric reaction systems as null-models to identify structural traces of evolution in metabolism

## Petter Holme

IceLab, Department of Physics,
Umeå University, 90187 Umeå, Sweden

Department of Energy Science,
Sungkyunkwan University, Suwon 440–746,
Korea

## Mikael Huss

Science for Life Laboratory Stockholm,
Box 1031, 171 21 Solna, Sweden

Department of Biochemistry and Biophysics,
Stockholm University, 106 91 Stockholm,
Sweden

## Sang Hoon Lee

IceLab, Department of Physics,
Umeå University, 90187 Umeå, Sweden

The metabolism is the motor behind the biological complexity of an organism. One problem of characterizing its large-scale structure is that it is hard to know what to compare it to. All chemical reaction systems are shaped by the same physics that gives molecules their stability and affinity to react. These fundamental factors cannot be captured by standard null-models based on randomization. The unique property of organismal metabolism is that it is controlled, to some extent, by an enzymatic machinery that is subject to evolution. In this paper, we explore the possibility that reaction systems of planetary atmospheres can serve as a null-model against which we can define metabolic structure and trace the influence of evolution.

We find that the two types of data can be distinguished by their respective degree distributions. This is especially clear when looking at the degree distribution of the reaction network (of reaction connected to each other if they involve the same molecular species). For the Earth's atmospheric network and the human metabolic network, we look into more detail for an underlying explanation of this deviation. However, we cannot pinpoint a single cause of the difference, rather there are several concurrent factors. By examining quantities relating to the modular-functional organization of the metabolism, we confirm that metabolic networks have a more complex modular organization than the atmospheric networks, but not much more.

We interpret the more variegated modular arrangement of metabolism as a trace of evolved functionality. On the other hand, it is quite remarkable how similar the structures of these two types of networks are, which emphasizes that the constraints from the chemical properties of the molecules has a larger influence in shaping the reaction system than does natural selection.

Reaction systems are, at many levels of the universe, motors driving the creation of higher structure. From the metabolism in our bodies, via reactions in planetary interiors and atmospheres, to the nuclear reaction systems in stars; these are all systems shaped by the physical properties of constituents—the atoms and molecules. Among these systems, metabolism is special in the sense that its control has evolved by natural selection. But the physical properties of molecules and the relative abundance of elements constrain the evolution of this genetic control. Perhaps these constraints explain that very different reaction systems—reactions in planetary atmospheres and the organismal metabolism—share large-scale features (like the right-skewed probability distributions of degree, which roughly speaking reflects the number of molecules a molecule can react with) [1,2]. Still, as we will see, there are differences between these two types of systems and in this paper we will focus on what these differences are and what they can tell us of the evolution of metabolism. To put it short, we explore the idea that the reaction systems of planetary atmospheres can be null-models for studying metabolic networks in an evolutionary perspective.

The study of reaction-system topology (the set of all participating reactions) has long been restricted, by lack of data, to small subsystems. These systems, like e.g. the citric acid cycle of metabolism [3] or the carbon-nitrogen-oxygen cycle of stellar nuclear reactions [4] (two systems that were, coincidentally, both discovered in the mid-1930's), have been modeled in great detail with e.g. differential equations. It has, however, not until recently been possible to investigate the system-wide organization of any type of reaction system. Since about a decade, we do have methods to infer the entire set of reactions (again coincidentally) both in metabolism and planetary atmospheres. Still these datasets are so crude that our conclusions in this paper will be rather hypothetical in nature. On the encouraging side, however, the early conclusions mentioned above—that reaction network are right-skewed and fat-tailed [1,2]—still hold for contemporary datasets. If we go beyond the topology, even less is known. A full picture of reaction rates and concentrations for a traditional kinetic modeling is far into the future. One complication comes from the fact that metabolites (and also molecular species in atmospheres) are distributed heterogeneously in space [5] and sometimes so few in number that concentration based models do not apply. This means that when investigating the global organization of reaction systems, we will have to rely on graph-based analysis techniques for still some time. Even though graph-based methods need to discard much of the knowledge we have about reaction kinetics, one can still encode much information into the graph. The molecular species present determine the vertices of the network; the catalysts present define the reactions. But what should the edges represent? Should one also include separate vertex-types for reactions and catalysts? The fundamental trade-off is between a graph representation including more information and a simpler representation that suits a larger variety of analysis methods. Much of the recent development in the graph structure of reaction systems has focused on either adapting analysis techniques to complex and informative graph representations [6–9], or to find simple graph representations encoding as much relevant information as possible [10–12]. In this paper, we will focus more on the latter developments and study the topology of two simple graph representations: one *substance graph* where the vertices are molecular species and an edge represents that two vertices participate in the same reaction, and a *reaction graph* where vertices symbolize reactions and two vertices are linked if they share some molecular species. In addition to these representations we also study the reaction systems as a bipartite graph with two classes of vertices, one for reactions and one for molecular species with edges connecting substances to the reactions they participate in. (Note that this representation, although more informative, still means a reduction of the information from the entire reaction system since one no longer can see which reactants that need to be present for a reaction to occur, or which products that are produced.) We investigate several topological properties of such graphs from reaction systems of planetary atmospheres and organismal data sets. Apart from degree distributions, we study network modularity (reflecting how well a graph can be decomposed into dense sub-graphs that are relatively weakly interconnected), currency substances (abundant molecular species that can react with a broad spectrum of other substances) and degree correlations (if edges primarily go between vertices of similar degree, or if the degrees are unbalanced with many edges between high- and low-degree vertices).

## The different degree distributions of the human metabolic and Earth atmospheric networks

Since the degree of a vertex count the number of other vertices it interacts with, it is a fundamental network quantity. The high-degree vertices can, and in most situations will, interact with many other vertices. The early findings that reaction systems have fat-tailed degree distributions—i.e. most vertices interacts only with a few others while some interact with a number far larger than the average—points at a diversity of functions among the vertices. For the metabolism, the common interpretation is that the high-degree metabolites are supplying building blocks to metabolites with more specialized functions, and lower degree. For atmospheric reaction networks, the low-degree vertices typically correspond to more complex molecules. We start our comparison of planetary and metabolic reaction system by looking at the substance and reaction graphs of Earth's atmosphere and the human metabolism. In Fig. 1A, we show the degree distributions of the substance graphs of the human metabolism and Earth's atmospheric reaction system. These distributions are rather similar—peaked and right skewed with tails of about the same slope. The degree distributions of the reaction graph, seen in Fig. 1B, are strikingly different. The human reaction graph is skewed and fat-tailed like its substance graph (but with a smaller exponent), whereas the Earth reaction graph has a degree distribution of an entirely different functional form, suggesting a different organization. The graphs are too big, however, for layout programs to give a hint of a deeper explanation of this difference (Fig. 2). Indeed, it is difficult to single out a more fundamental quantity causing the differences in degree distributions, as we will see in the rest of this section.

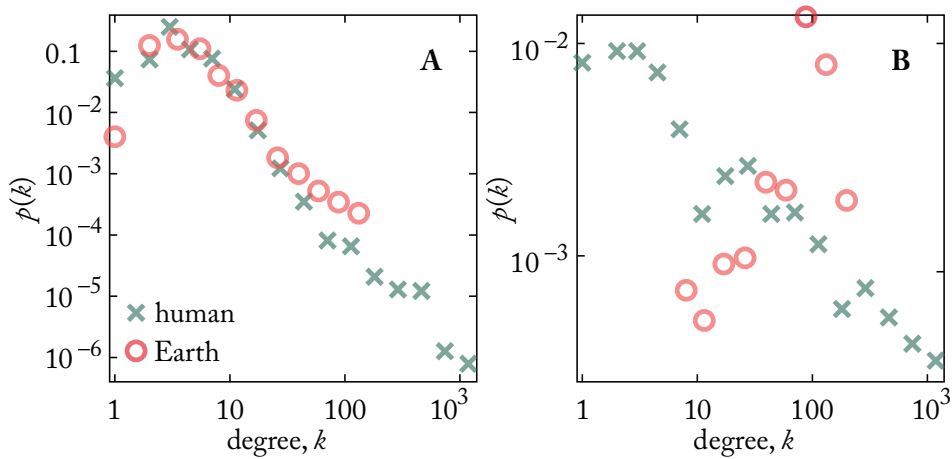

*Figure 1. Degree distributions of substance and reaction graphs of the human metabolism and Earth's atmospheric reaction system.* Panel A shows the probability mass-function of the degree of the substance graph of the reaction system of the Earth's atmosphere and the human metabolic networks. B shows the same as A, but for the reaction network. The similar behavior in A is drastically different in B. The plots are log-binned and plotted on double logarithmic scales.

In our quest for a more detailed explanation of the difference of degree distributions in Fig. 1, we look closer at the bipartite representations mentioned above. In Fig. 3 (panels A, B, E and F), we plot the probability distribution of bipartite degree $K_i$ for the human metabolic (Figs. 3A and B) and Earth atmospheric (Figs. 3E and F) networks in the substance (Figs. 3A and E) and reaction (Figs. 3B and F) projections. (For the other data sets this information can be found in the Supporting Information.) For substances, the degree distributions are right skewed in a fashion similar to the substance graph of Fig. 1A. For reactions, the two types of reaction systems both show unimodal degree distributions. A slight difference is that the Earth data gives a left-skewed distribution while the human network is right-skewed. This also means that the bipartite reaction-degree distribution, for the human data, is radically different than the projected distribution of Fig. 1B. To understand this better, we can decompose the degrees of the projected networks into three quantities as follows (where the left-hand side is the degree of the projected network and the right-hand side quantities refer to the bipartite representations):

$$k_i = S_i - K_i - X_i = K_i(\kappa_i - 1) - X_i, \quad (1)$$

where $S_i$ is the sum of degrees of $i$'s neighbors, $K_i$ is $i$'s degree, $X_i$ is the number of four-cycles that $i$ is a part of, and $\kappa_i$ is the average degree of $i$'s neighbors. If there are few four-cycles in the bipartite network and there are no strong degree correlations (so $\kappa_i$ can be assumed constant with respect to $k_i$), then $i$'s degree in the bipartite network is a linear function of $k_i$ (according to Eq. (1)). This is thus not the case for, at least, the metabolic reaction network where the $k$- and $K$-degree distribution, as mentioned, differs much. Indeed, in Fig. 3B we see a positive correlation between $K$ and $\kappa$, stronger than the corresponding correlation for the Earth network in Fig. 3F (which is almost absent). This means that $S = K\kappa$ grows superlinearly with $K$ so the tail of $p(K)$ gets stretched into the distribution of Fig. 1B. Here, we still assume that the number of four-cycles does not contribute to $k$ significantly, which we justify below. This is justified to some extent in Fig. 3C (and 3G for the Earth network)—the $k$-scaling of $S$ and $X$ is similar, so $S - X$ scales like $S$ (and thus the arguments above still hold). That $S$ (and thus $S - X$) scales like $X$ is also true for the atmospheric network (Fig. 3G), which explains that the shape of Fig. 1B is to a large degree determined by $K$ (so the hump shape of $p(K)$ gives a hump-shaped $p(k)$). Another view of $S$, $K$ and $X$ is given in panels D and H where, we plot the average degrees of nodes given their $S$-, $K$- and $X$-values. We can see that, as expected, $S$ is the best predictor of (showing close to a linear relationships for the metabolic data, and a clear correlation for the atmospheric network). Another observation is that $X$ shows more structure (apart from the scaling itself) in the metabolic network compared to the atmospheric network. This can perhaps be explained by the more pronounced modular structure of the metabolic network (that we will discuss further below). From Fig. 3D and H we also learn that shows a strong positive $K$-dependence for the metabolic network, but not for Earth's atmospheric network. This is reflected in Figs. 3B and F too—since κ grows with $K$ for the metabolic network, $S$ and $K$ will be positively correlated, and since grows with $S$ then it will also grow with $K$.

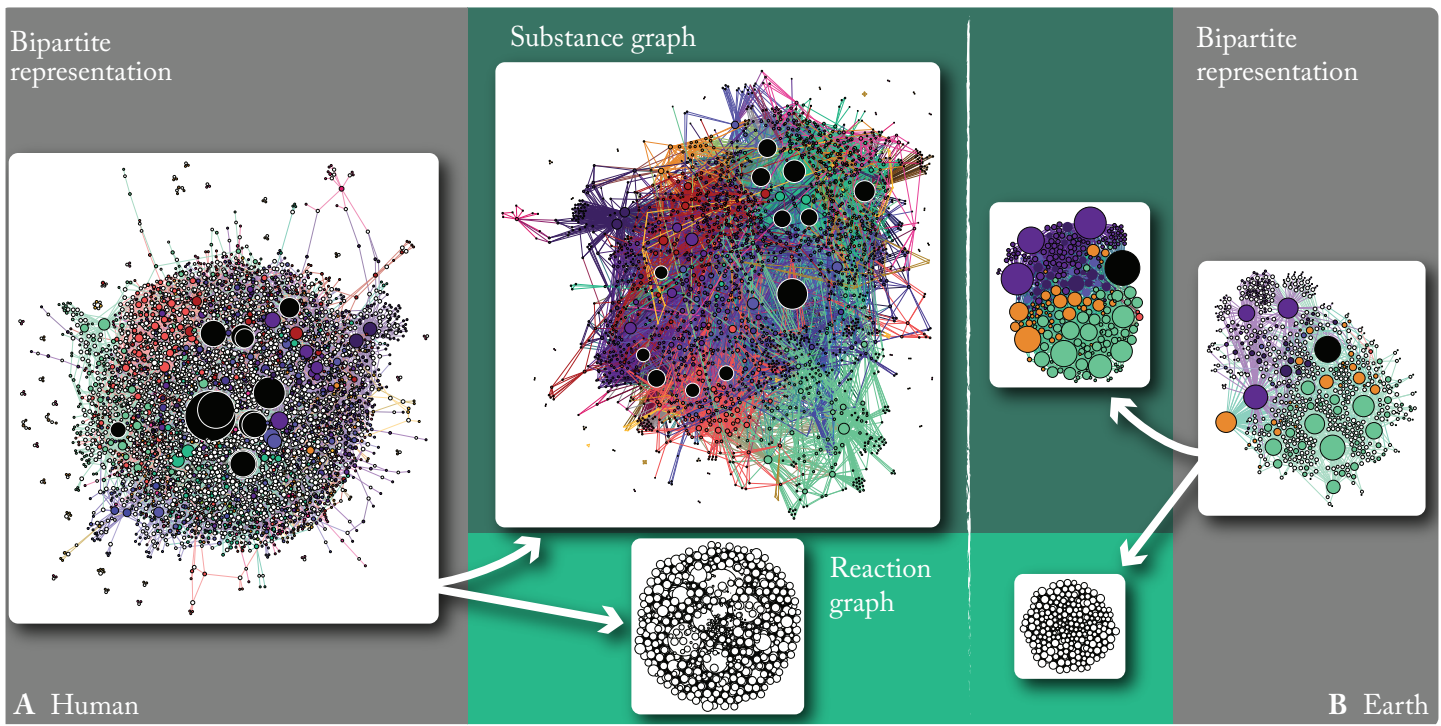

*Figure 2. Ridiculograms of the human metabolism and Earth's atmospheric reaction system in bipartite, substance and reaction graph representations. The areas of the vertices are proportional to their degree. White vertices are reaction vertices; black vertices are currency vertices. For the other vertices the color represent different network modules. The colors of the edges are the same as their vertex of largest degree.*

In summary, the difference between the degree distributions of the reaction graphs of the metabolic and atmospheric networks cannot be explained by one single feature of the original reaction system's topology. Instead it can be traced to a combination of the slightly different skewness of the distribution of a reaction's number of participating substances and the different correlation properties between the degree of a vertex and the average degree of its neighbors. In the Supporting Information, we plot the bipartite degree distributions of all the planets and organisms. Essentially, the conclusions for the Earth's atmospheric network extends to other planets, except that the data sets are smaller and the degree distributions does not have the same negative trend similar to power-laws.

**Comparing degree distributions of planetary atmospheric and organismal metabolic networks**

So far, we focused on finding lower-level causes for the degree distributions of projected networks of the human metabolic and Earth atmospheric networks. We now turn to the question how much these observations can be generalized to the other networks. To this end, we will use more rigorous methods for analyzing probability distributions than we used so far. We will analyze the data using methods from Ref. [11]. First, we test the hypothesis that degrees are power-law distributed by (roughly speaking, details in the Methods section) finding parameter values for the power-law distribution that fits the data best, then draw as many series of numbers from this distribution with the same size as the raw data and check the likelihood that the synthetic and real data come from the same distribution. We also check which is the most likely distribution generating that degree distribution—power-law or log-normal (a right-skewed distribution with a more narrow tail than a power-law that is visually similar to the Earth reaction graph of Figure. 1B). The results of these measurements are shown in Table 1. As hypothesized above, the reaction graphs are unanimously inconsistent with power-laws. Of the substance graphs, only planetary atmospheric networks are consistent with power-laws. This does not mean that it is fair to describe them as power-laws; especially since most of them fit better to a log-normal form. Since the planetary data sets are relatively small, the relative errors are larger and it is harder to refute the possibility of another functional form. The substance graphs are, on the other hand, closer to log-normals than power-laws. The reason is seen for the human metabolic network in Fig. 1 (and for the other data sets in the Supporting Information), that they are even more fat-tailed than a power-law—they have more vertices of highest degrees than the best-fitting power-law does. Thus they are even further from log-normals than power-laws. The reaction graphs are more similar to power-laws than log-normals for the metabolic networks, but the other way around for the planetary atmospheres, which is also in line with our observations. This study cannot, however, strengthen the observation that the substance graphs are similar to the metabolic networks except for Earth's network that falls into the same category as the metabolic networks. There are two possibilities—either the difference can be explained by a difference in sizes and that the other planetary atmospheres have to be measured by indirect methods, or the Earth network is radically different (more than just the sizes). Ref. [2] makes the latter hypothesis, and argues a difference from the influence on the biosphere on the Earth's atmosphere creates a visible difference. On the other hand, many reactions typical for Earth (e.g. involving molecular oxygen) are also present in the other datasets.

The substances' degrees in the bipartite representation do not separate the planetary and metabolic data so well (both types of datasets contain degree distributions consistent with power laws, and not). Similar to the observations in the detailed studies above, the projections to substance or reaction graphs create the difference However, the planet-network distributions are more similar to log-normal than power-laws, whereas it is the other way around for the metabolic networks.

## Modularity and currency metabolites

Biological systems are commonly described as modular—being composed of different subunits, or modules, which perform some specific task relatively independent of the rest of the system. Some modules are quite conspicuous—a cell is a prime example—but also more nebulous systems, like metabolism, are thought to consist of modules. If we treat all reactions equal (the essence of the graph theoretic approach), then independence means that the connections within the network module should be denser than the connections out of the module. A module on a graph-level resolution of metabolism is thus equal to what is commonly known as a network cluster or community [13]. This is not quite the whole story however. The most abundant metabolites (like water, carbon dioxide and so on) do not put any restriction on the reactions, and would not contribute to the specialized function of a module. It is thus common to preprocess the graph by identifying such *currency metabolites* and removing them from the network, considering only a network of other less frequent molecular species that are more of bottlenecks in the metabolic machinery. There are methods to identify both network clusters and currency metabolites (described in the Methods section) from the topology of substance graphs. Although these definitions have been developed for metabolic networks, there is nothing that stops us from applying them to networks of planetary atmospheres. *A priori*, since atmospheric reaction system has not evolved through natural selection, we expect them to have less distinct modules and currency metabolites. This is indeed the case as can be seen in Fig. 4—there is a size-difference between the metabolic and atmospheric networks, but it is less pronounced than both the relative modularity and the number of currency vertices. Thus there seems to be a stronger tendency for the metabolic networks to be organized into modules supplied by currency vertices than the networks of planetary atmospheres.

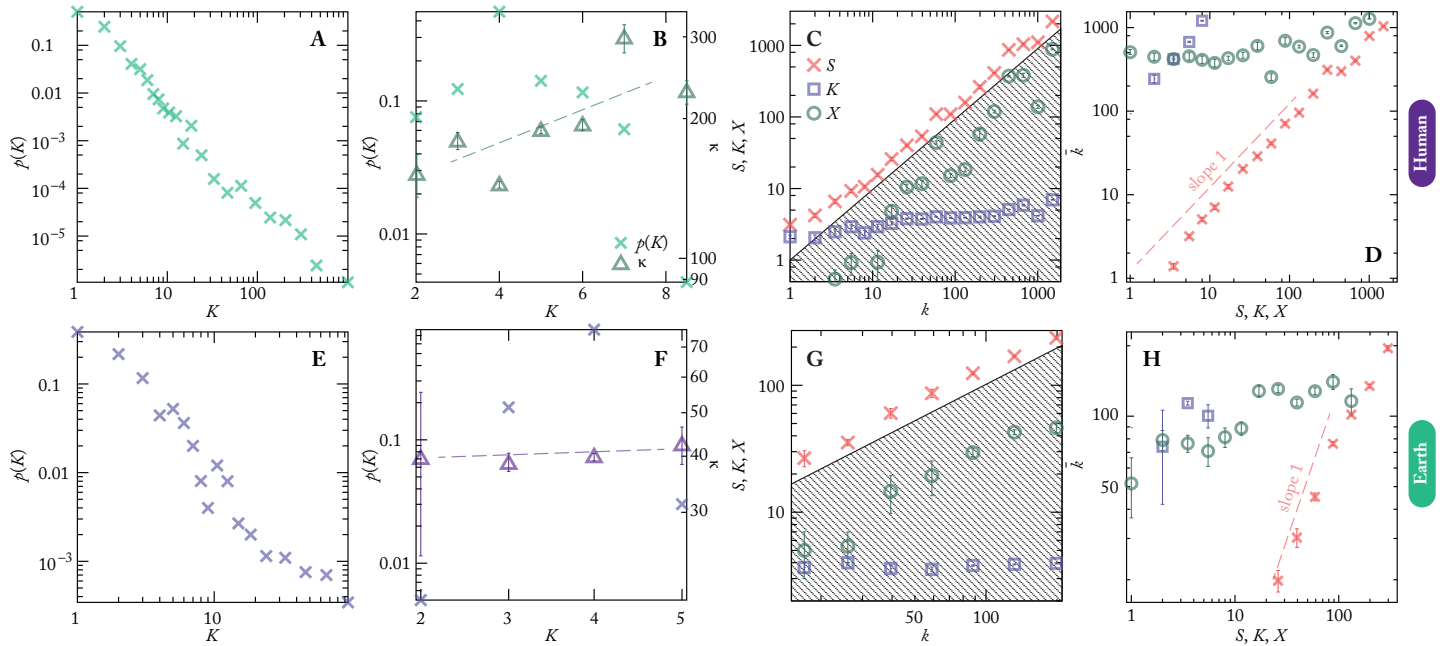

*Figure 3. Deeper investigations of the degree distributions.* Panel A displays the degree distribution of substances in a bipartite representation of the reaction system, i.e. the probability distribution of the number of reactions a substance participates in. Panel B shows the corresponding plot for reactions and also the average degree of neighbors. The dashed line is a linear-regression line to highlight the trend in κ. C and G displays the values of the three bipartite-network terms of $k$—$S$ (the sum of the degrees of neighbors), $K$ (the degree) and $X$ (the number of four-cycles the vertex participates in). The diagonal line shows the $k$-value (so if you subtract the values of circles and squares from the values of crosses you would get this line). Panel D and H shows the average degrees $\bar{k}$ of nodes with certain values of the three terms that contribute to the degree in the projected reaction networks. $\bar{k}$ is averaged over logarithmic bins of $S$, $K$, and $X$ values. The dashed line is a reference corresponding to a linear $\bar{k}$-dependence. Panels A–D are for the human metabolic reaction networks, E–H show the corresponding plots for the Earth atmospheric reaction networks.

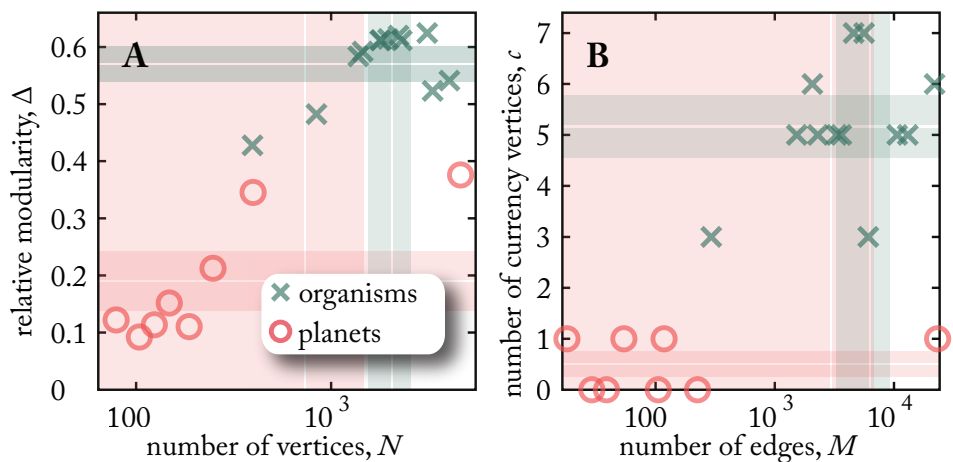

*Figure 4. Relative modularity and the number of currency vertices separate networks of metabolism from networks of planetary atmospheres more than their sizes do.* To show that the maximal relative modularity separates metabolism from reaction systems of planetary atmospheres, we display (panel A) the relative modularity $\Delta$ as a function of the number of vertices $N$. The shaded areas indicate the standard deviation and means of the respective quantities. Similarly, in B, we show another quantity related to the functional organization, the number of currency vertices $c$, as a function of the number of edges $M$ in the network. Note that axes are linear and logarithmic respectively.

## Discussion

In this article, we have directly compared functionally informative network characteristics of metabolic reaction systems of a wide variety of organisms and the reaction systems of planets and moons of the solar system. One such quantity is degree—the number of other nodes a node interacts with. (Where "interact" is defined via the network in question.) In most types of networks, degree indicates the importance of a node, but in biochemical networks, where both low- and high-degree vertices can be essential for the cell's functionality, then degree rather separates chemical substances of different functionality—at least in metabolic substance networks, the high-degree vertices are typically light molecules that supply atoms and molecular groups to the functionally more specialized low-degree vertices [14]. For reaction networks one can assume a similar interpretation—high-degree vertices are reactions supporting many subsystems of the reaction system. All substance projections, for both atmospheric and metabolic networks, do indeed have relatively broad degree distributions. This supports the above-mentioned picture of functional differentiation by degree. Using statistical tests, we can separate organisms from planet fairly well. The networks of planetary atmospheres are typically consistent with power-laws, but the metabolic networks are not. The planetary networks are, however, statistically more similar to log-normal distributions, which suggests that the fact they are deemed consistent with power-laws is an effect that they are, on average, smaller than the metabolic systems (and thus does not provide enough data to give statistical significance).

We note that in the substance-network projection, the Earth atmospheric and human metabolic networks have rather similar degree distributions, but for the reaction-network projection the distributions are strikingly different. We investigate lower-level explanations for this observation in terms of degree distributions of a bipartite representation of the reaction system and degree correlations. It is however not easy to single out a low-level cause for this difference, rather it seems to be a combined effect of a slightly difference in the distribution of reaction-degrees and degree correlations in the bipartite representation.

When we look closer at quantities designed to characterize the modular functionality, we see higher network modularity and more currency metabolites in metabolic networks than atmospheric networks. On the other hand, the differences are not larger than that they can almost be explained by the sizes of the networks alone. Furthermore, fundamental structures such as the shape of some of the degree distributions are skewed in a qualitatively similar way. Our conclusion is thus that the main structure of metabolic networks is probably shaped by the same fundamental stoichiometric constraints as all chemical reaction systems, but there are also traces of evolution in the network structure of metabolism. At the same time the network-modular structure, the traces of evolution, is not so clear as the picture the analogy to engineering paints—there are more than a couple of in- and output terminals. Maybe the largest open question is not why metabolic networks are modular but why they are not more modular? How can we reconcile the logical picture of evolution operating by adding and deleting of modules with the modular-but-not-very-much-so picture of metabolic networks? We believe the approach we take in this paper, to use a natural system as a null-model for the metabolism can be fruitful.

| | | Atmospheres of planets and moons | | | | | | | | Metabolism of organisms | | | | | | | | |
|---|---|---|---|---|---|---|---|---|---|---|---|---|---|---|---|---|---|---|
| | | Earth | Venus | Titan | Titan 2 | Mars | Jupiter | Io | Solar system | Human (KEGG) | Human (BiGG) | M. genitalium | S. cerevisiae (KEGG) | S. cerevisiae (BiGG) | E. coli | M. musculus | D. melanogaster | C. elegans |
| Substance graph | Power-law? | Y | N | Y | Y | Y | Y | Y | Y | N | N | N | N | N | N | N | N | N |
| | PL or LN | PL | LN | LN | LN | LN | LN | LN | LN | PL | PL | PL | PL | PL | PL | PL | PL | PL |
| Reaction Graph | Power-law? | N | N | N | N | N | N | N | N | N | N | N | N | N | N | N | N | N |
| | PL or LN | LN | LN | LN | LN | LN | LN | LN | LN | PL | LN | PL | PL | PL | PL | PL | PL | PL |
| Bipartite substances | Power-law? | Y | Y | Y | N | N | Y | Y | Y | N | Y | Y | N | Y | N | N | N | Y |
| | PL or LN | PL | LN | LN | LN | LN | LN | LN | LN | PL | PL | PL | PL | PL | PL | PL | PL | PL |

Table 1. **Statistical tests of various types of degree distributions.** Statistics of the reactions in the bipartite representation are omitted since they are not fat-tailed. "Y" ("N") indicates that the data set is consistent (inconsistent) with the tested hypothesis. "PL" stands for "power-law" (i.e., testing for a power-law hypothesis); "LN" means "log-normal".

## METHODS

### Datasets for metabolic and chemical networks

Reaction sets for planetary atmospheres are described in Ref. [5], except the "solar system" data that was obtained from the UMIST database [15]. The metabolic networks come from the KEGG [16] and BiGG [17] database and are described in Ref. [7]. We select nine datasets from the KEGG and BiGG databases to match the number of planetary atmosphere datasets. To get a rough error estimate of sampling effects, we also analyze the human data both from BiGG and KEGG, and two independent datasets from Jupiter's atmosphere. Our selection criterion is that the datasets should be a diverse selection among the most well-studied model organisms.

### Network representations

To choose the graph representation of a reaction system involves a trade-off between information content and usefulness. One can use a complex representation with substances, catalysts and reactions as separate classes of vertices and directed edges representing the general direction of the matter flow. The advantage with such a representation is that all topological aspects of the reaction system are encoded into the graph. But the price for this is that there few general analysis methods can be applied to it; they would need to be modified, something that is not always possible. Alternatively, one chooses a simple-graph representation with one type of vertices and one type of (undirected) edges, without multiple edges or self-edges. Such a representation can be analyzed by a multitude of off-the-shelf methods. A disadvantage with simple graphs, except that they encode less information, is that there is no obvious way of reducing the reaction system to a simple graph. We choose a substance graphs as our main graph simple-graph representation. In such a graph one put an edge between all substances that can participate in the same reaction, so if the reaction $2H_2O \to 2H_2 + O_2$, would contribute with three edges—$(H_2O,H_2)$, $(H_2,O_2)$ and $(O_2,H_2O)$—to a substance graph. There is some evidence that substance graphs are good simple-graph representations of metabolic networks [11,18], but to the best of our knowledge, no corresponding studies for other categories of reaction systems. In addition, we use a reaction graph representation that is in some sense dual to the substance graphs—every reaction is a vertex in this network and two reactions that have a substance in common is connected.

### Testing degree distributions

We use the approach in Clauset *et al.* [11] to test the degree distributions for the hypothesis that they follow power-laws. This method starts from the real data and obtains the exponent of a best-fitting power-law, α, by maximum likelihood estimation. Then one draws sets of random numbers, of the same cardinality as the original data, from the probability distribution

$$p_k = \begin{cases} \Lambda k^{-\alpha} & \text{if } 0 < k \leq k_{\max} \\ 0 & \text{otherwise} \end{cases} \quad (2)$$

where $\Lambda$ is a normalization constant. Finally, one use the Kolmogorov–Smirnov test statistics (the maximal difference,

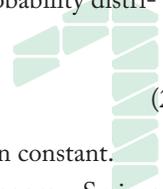

for all $k$-values, between the cumulative density functions of the real and synthetic data) to estimate the p-value of the hypothesis that the real data was drawn from $p_k$.

Ref. [19] also adapts a method by Vuong [20] to compare different heavy-tailed distributions. We use it to test which distribution of power-law and log-normal distribution functions that best fits our data. The log-normal distribution is defined by the probability density function

$$p_k = \frac{A'}{k}\exp(-a'(\ln k - \mu)^2) \quad (3)$$

where $A'$, $a'$ and $\mu$ are positive constants ($A'$ is a normalization factor, $a'$ and $\mu$ are parameters giving the shape of the curve). Vuong's method takes the likelihoods, $L_1$ and $L_2$, of the two functional forms generating the observed data as its starting point. The method uses the result that is normally distributed for large data sets to compute a p-value for the hypothesis that the data was generated by distribution 1 rather than distribution 2.

**Network modularity**

The concept of network modularity, cluster, or community structure strives to capture the large-scale organization of networks into dense subnetworks that are relatively weakly interconnected [21]. There is no unique way of deriving a measure for network modularity or dividing a graph into such dense subgraphs; rather, there is a number of different methods each capturing some certain aspect of network modularity. The method in this work is based on the popular method of maximizing Newman and Girvan's $Q$-modularity. For this measure, one assume the graph is divided into a number of subgraphs and let $e_{ij}$ be the fraction of all edges going between subgraph $i$ and $j$, and defines

$$Q = \sum_i (e_{ii} - (\sum_j e_{ij})^2) \quad (4)$$

A class of module-detection methods starts by assuming that the division maximizing $Q$ is a sensible decomposition into subgraphs. Already from the Equation (4) one can see that edges within a subgraph give a positive contribution to $Q$, and edges between communities decrease $Q$. The advantages with this clustering algorithm are that $Q$ is easy to interpret and closely matching the verbal definition of a network module above; and furthermore the maximal $Q$, $Q'$, is a crude measure of the network modularity of an entire graph. The two disadvantages with $Q$-maximization methods are the following. First, it fails to divide some subgraphs into what looks like obvious clusters. This is roughly speaking because the second sum compares a division $i$ with all other divisions $j$, even if it does not matter (for a visually good clustering) if $i$ and $j$ are far apart [22]. Second, it is technically hard to find the maximizing division—$Q$ is a very flat function (in sub-division space) near its maximum [23]. For our purpose these latter two objections are not so serious—there is no general biological argument that the modules that look like they can be further subdivided are not sensible clusters, and there is no need to find the actual subdivision into modules, we just want a good estimate of $Q'$, which we do have if we only get close to the mentioned plateau in subdivision space.

As a measure of the modularity of a graph, $Q'$, is not ideal. On one hand $Q'$ close to zero would mean a low modularity and $Q'$ close to one would imply modularity. On the other hand, the intermediate values depend on many factors regarded as more fundamental (like the number of vertices and edges and the degree distribution) than modularity. To compensate for such effects as much as possible we rather measure $Q'$ relative to the average value of $Q'$ in an ensemble, or null-model, of graphs (obtained by standard edge rewiring [24]) with the same sizes $N$ and $M$ and the same degrees as the substance graph $G$, but everything else random. So we define

$$\Delta = Q' - \overline{Q} \quad (5)$$

where $Q'$ is the average of the maximal modularity over 1000 rewired graphs.

**Currency vertices**

The hubs in metabolic networks—e.g. $H_2O$, NADH, ATP and $CO_2$—are typically also the most abundant metabolites throughout the cell. These are the workhorses of metabolism, supplying functional groups to proteins and other molecules with more specialized functions. Since these currency metabolites are present throughout the cell and do not put much of constraints on the reactions they participate in, one can learn more about the functionality of the network if one exclude them from the graph representation. The circumstance that they are common throughout the cell and participate in many reactions also means that they connect network modules and effectively lower the modularity. This observation, along with the fact they have a high degree, has been used as a definition of currency metabolites [10]. If one deletes vertices in order of their degree (starting from large degrees) and monitor $\Delta$, then for metabolic networks, $\Delta$ typically first increase to a maximum and later decrease. Ref. [10] defines currency metabolites as those that give the largest $\Delta$ before $\Delta$ reached a value larger than in the original graph. This definition is general enough to apply to other reaction-system networks, and one can speak of *currency vertices* also for atmospheric or nuclear reaction systems [14].

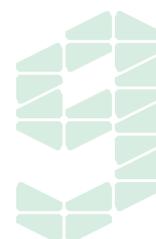

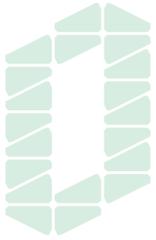
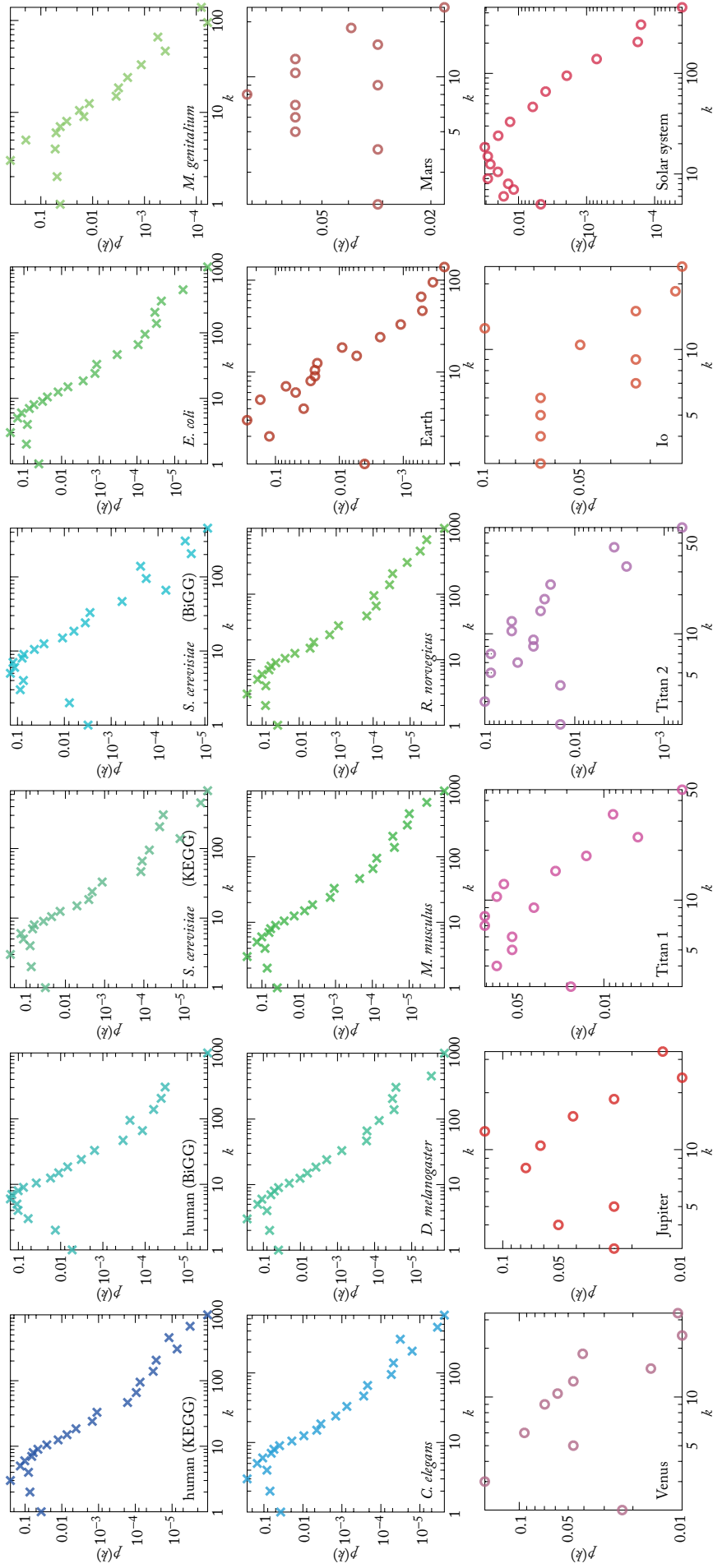

*Fig. S1. Degree distributions for the substance networks. The data is log-binned and plotted in log–log scale.*

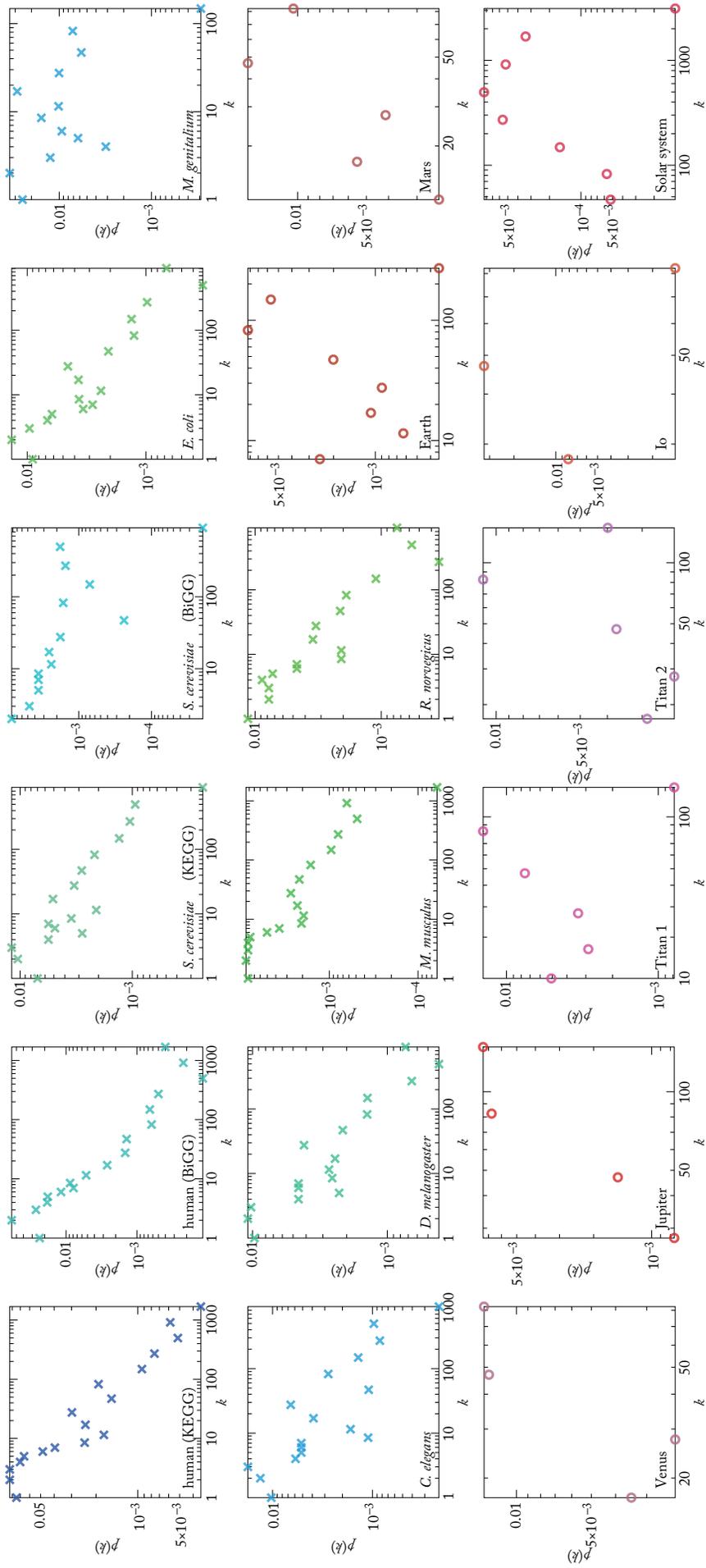
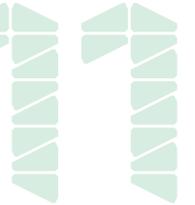

*Fig. S2. Degree distributions for the reaction networks. The data is log–binned and plotted in log–log scale.*

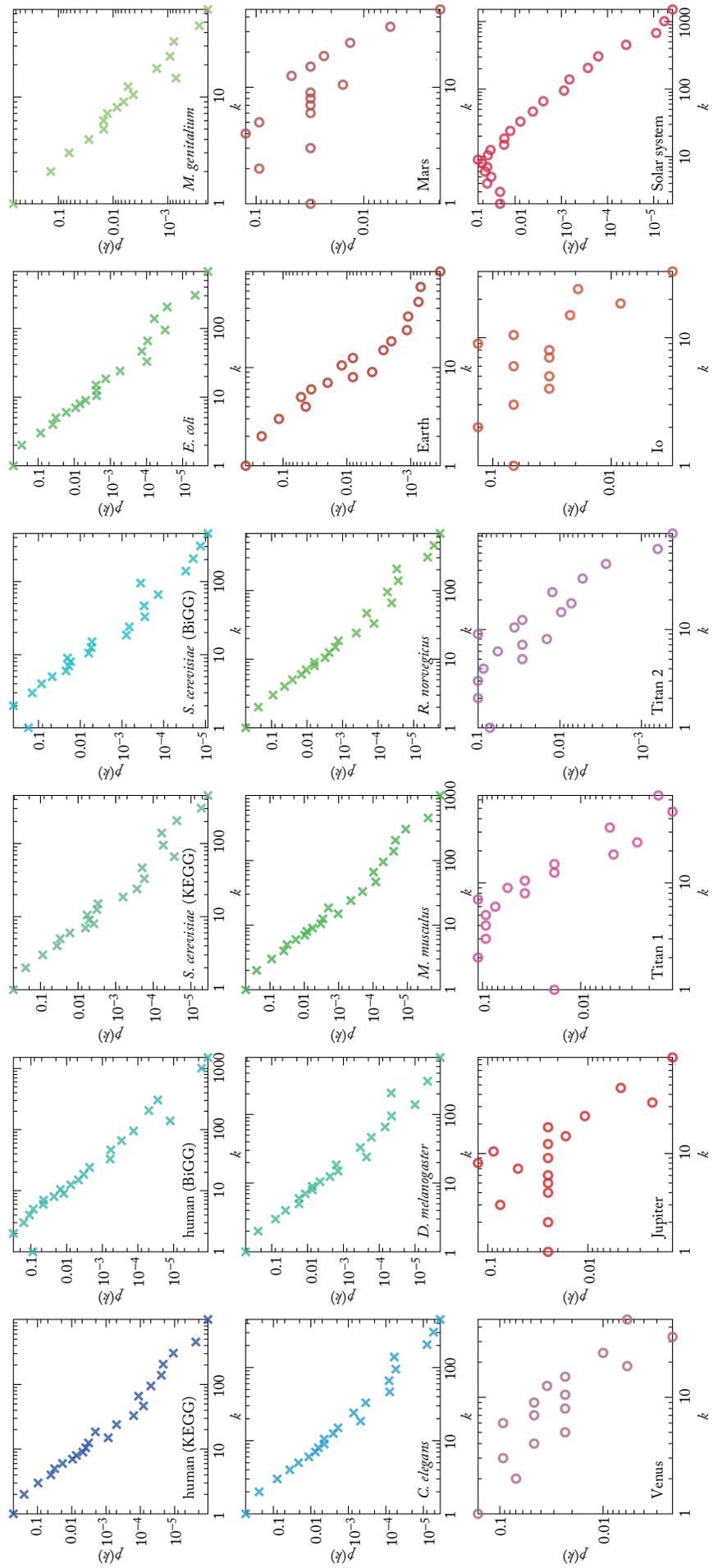

**Fig. S3. Degree distributions for the substances in the bipartite representations.** *The data is log-binned and plotted in log–log scale.*

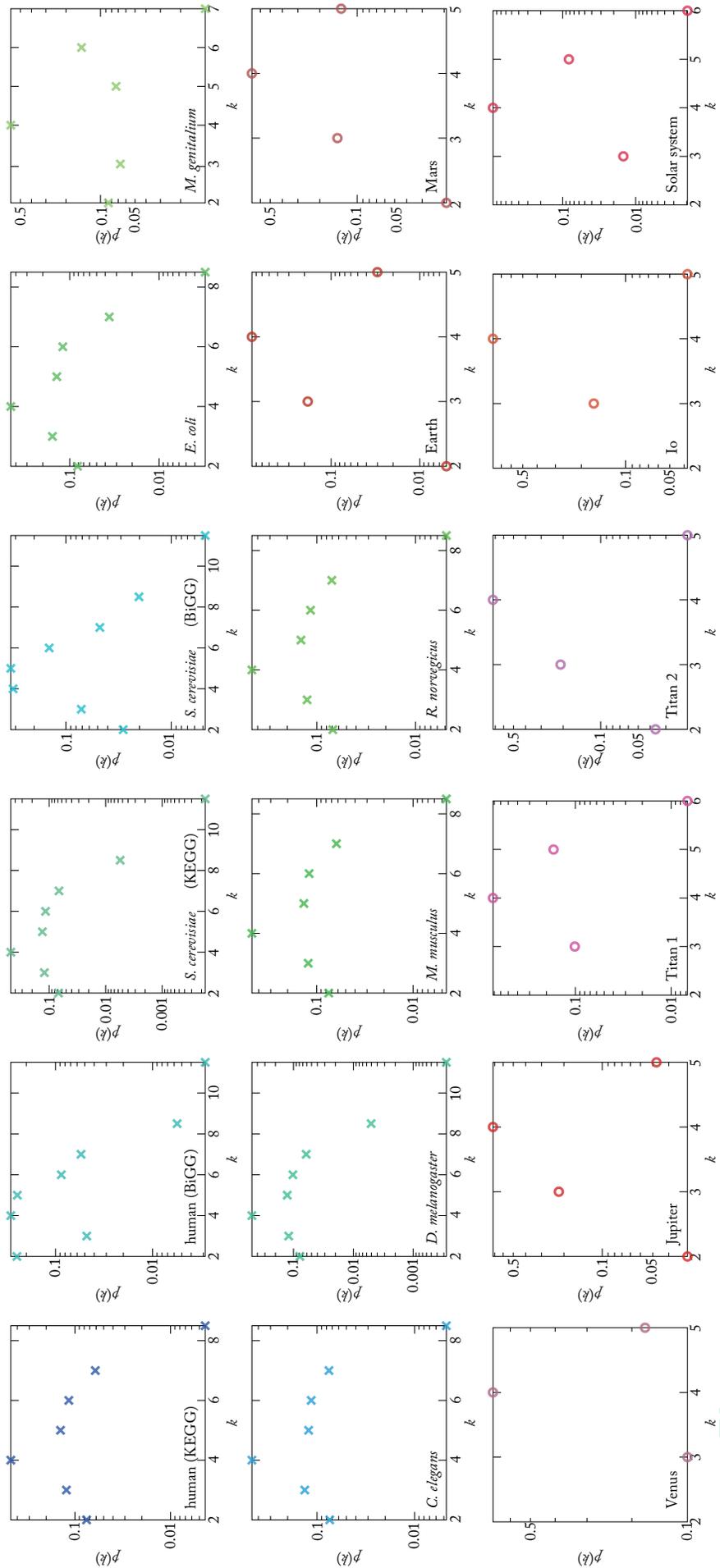
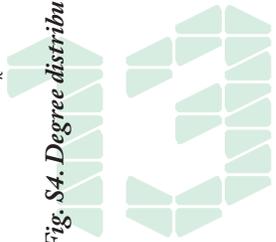

*Fig. S4. Degree distributions for the reactions in the bipartite representations. The data is log-binned and plotted in log-log scale.*

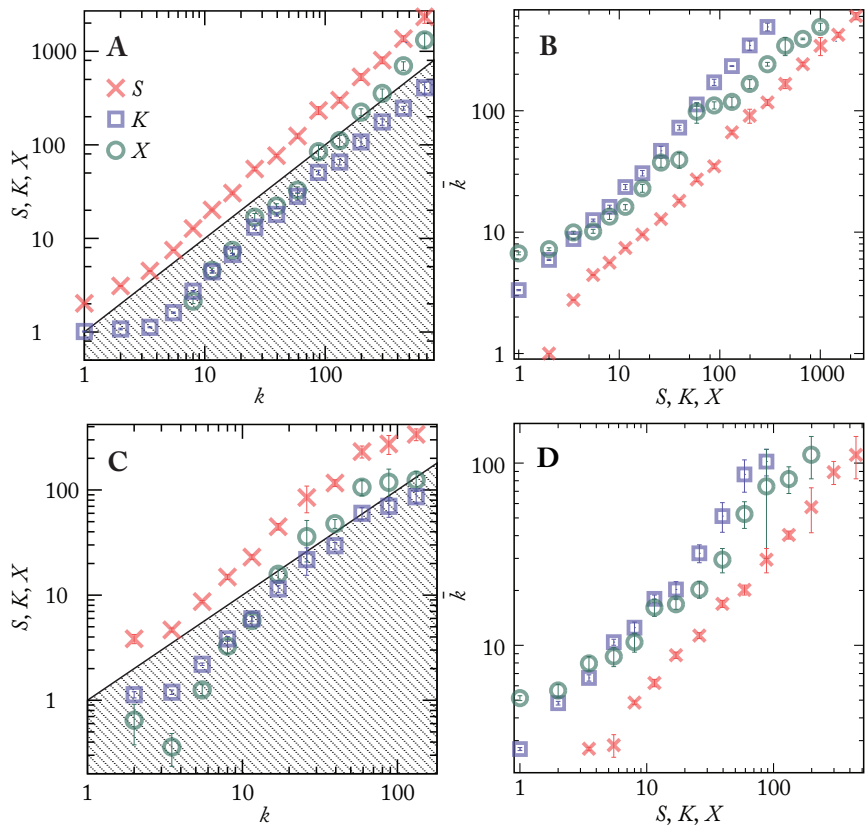

*Fig. S5. A plot corresponding to Fig. 3C, D, G and H for substance networks.* Panels A and C display the values of the three terms of k—S, K and X. The diagonal line shows the k-value. Panels B and D show the average degrees $\bar{k}$ of nodes with certain values of the three terms that contribute to the degree in the projected networks. $\bar{k}$ is averaged over logarithmic bins of S, K, and X values. Panels A and B is data for the human network; C and D are the corresponding plots for the Earth atmospheric network.

Human

Earth

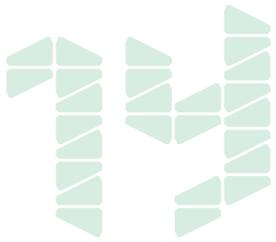